\documentclass[prc,showpacs,showkeys,superscriptaddress,nofootinbib,twocolumn,floatfix]{revtex4}
\usepackage{graphicx,color,mathrsfs,amsmath,amssymb,amsthm,amsopn,mathbbol,bm} 
\newlength{\lslashl}
\def\FMslash#1{\settowidth{\lslashl}{$#1$}\makebox[\lslashl]{\makebox[0mm]{$/$}\makebox[0mm]{$#1$}}}

\def\kslash{\FMslash k}
\def\rslash{\FMslash r}
\def\vec#1{\mathbf #1}

\def\ie{{\em i.e.}}
\begin{document}

\title{Selfconsistent Evaluation of Charm and Charmonium in the 
Quark-Gluon Plasma} 

\author{F. Riek}\thanks{e-mail: friek@comp.tamu.edu}
\affiliation{Cyclotron Institute, Texas A\&M University, 
College Station, Texas 77843-3366, USA}
\author{R. Rapp}\thanks{e-mail: rapp@comp.tamu.edu}
\affiliation{Cyclotron Institute, Texas A\&M University, 
College Station, Texas 77843-3366, USA}

\begin{abstract}
A selfconsistent calculation of heavy-quark (HQ) and quarkonium 
properties in the Quark-Gluon Plasma (QGP) is conducted to quantify 
flavor transport and color screening in the medium. The main tool 
is a thermodynamic $T$-matrix approach to compute HQ and quarkonium 
spectral functions in both scattering and bound-state regimes. The 
$T$-matrix, in turn, is employed to calculate HQ selfenergies which 
are implemented into spectral functions beyond the quasiparticle 
approximation. Charmonium spectral functions are used to evaluate 
eulcidean-time correlation functions which are compared to results 
from thermal lattice QCD. The comparisons are performed in various 
hadronic channels including zero-mode contributions consistently 
accounting for finite charm-quark width effects. The zero modes are 
closely related to the charm-quark number susceptibility which is
also compared to existing lattice ``data". Both the susceptibility and 
the heavy-light quark $T$-matrix are applied to calculate the thermal 
charm-quark relaxation rate, or, equivalently, the charm diffusion 
constant in the QGP. Implications of our findings in the HQ sector for 
the viscosity-to-entropy-density ratio of the QGP are briefly 
discussed. 
\end{abstract}

\date{\today}
\maketitle
%
\section{Introduction}
In recent years rather remarkable properties of the Quark-Gluon Plasma
(QGP) have been discovered, notably a liquid-like behavior deduced
from the collective expansion of the medium formed in collisions of
heavy nuclei at the Relativistic Heavy Ion Collider (RHIC). Hydrodynamic
expansion models, based on the assumption of local thermal equilibrium,
require very short equilibration times, of the order of 
$\tau_{\rm therm}\simeq1$\,fm/$c$ or less, to account for the 
experimental findings. The microscopic origin of the rapid thermalization 
remains one of the outstanding problems raised by the RHIC data. It is 
further complicated by the fact that the very notion of equilibrium 
erases information about its origin. A phenomenologically suitable probe 
of thermalization in heavy-ion collisions should therefore interact 
strongly at the thermal scale but with a relaxation time on the order 
of the system's lifetime (at RHIC $\tau_{\rm QGP}\simeq5$\,fm/$c$). 
The logical choice are thus 
heavy quarks, i.e., charm and bottom ($Q=c,b$), for which thermalization 
is expected to be delayed by a factor $m_Q/T$, quite comparable to 
$\tau_{\rm QGP}/\tau_{\rm therm}$ (see, e.g., Ref.~\cite{Rapp:2009my} 
for a recent review).

Bound states of heavy quarks (charmonia and bottomonia) are believed to 
probe the QGP from a somewhat different 
angle~\cite{Satz:2005hx,Rapp:2008tf,Bazavov:2009us,BraunMunzinger:2009ih,Kluberg:2009wc}.
The dissolution pattern of the quarkonium spectrum in matter is possibly 
the most direct way of diagnosing color-Debye screening of the basic QCD
force, $\vec f_{Q\bar Q} = -\vec\nabla V_{\bar QQ}(r)$, as a function of 
temperature (and/or density). In-medium quarkonium spectroscopy therefore
reveals insights into the deconfining hadron-to-parton phase transition in 
QCD. In particular, the use of potential models has been revived recently, 
largely triggered by the prospect that an in-medium potential can be 
extracted model-independently from lattice QCD (lQCD) computations. 
Furthermore, lQCD generates euclidean-time ($\tau$) correlation 
functions of quarkonia with good accuracy in different hadronic channels, 
which provide useful constraints on model calculations of spectral 
functions in the timelike regime of physical excitations. The extraction 
of reliable information from such comparisons requires a realistic 
modeling of the continuum part (scattering regime) of the quarkonium 
spectral functions, i.e., not only its bound-state part. The account of 
interactions in the near-threshold region is particularly important to 
describe situations were bound states dissolve into the continuum, as 
in the case at hand. In connection with correlator analyses, a 
comprehensive treatment of bound-state and continuum regimes has been 
performed using Schr\"odinger phase shifts~\cite{Wong:2006bx}, a 
thermodynamic $T$-matrix~\cite{Cabrera:2006wh} or a nonrelativistic 
Green's function~\cite{Mocsy:2007jz} approach.
Another important ingredient to a realistic description of quarkonium 
spectral functions in the QGP are medium effects in the single-particle
properties, i.e., in the heavy-quark (HQ) propagation. These are closely 
related to HQ transport~\cite{Petreczky:2005nh,Riek:2010fk} and the 
so-called zero-mode contributions to quarkonium spectral and correlation
functions~\cite{Aarts:2005hg,Umeda:2007hy}. In particular,
finite-width effects on both quarkonia and heavy quarks have received
little attention thus far~\cite{Cabrera:2006wh,vanHees:2007me}.    

In the present paper we further develop our previous study of heavy 
quarks and quarkonia in the QGP using a thermodynamic 
$T$-matrix approach~\cite{Riek:2010fk}. First, we go beyond the 
quasi-particle approximation for heavy quarks and perform 
a selfconsistent calculation of the HQ selfenergy and the in-medium 
heavy-light $T$-Matrix. The improved HQ spectral functions are then 
implemented into an off-shell calculation of the quarkonium spectral 
functions. Second, we expand our comparison with Euclidean correlator 
ratios to the scalar, vector and axialvector channels. In these channels
the presence of zero modes~\cite{Aarts:2005hg,Umeda:2007hy}, 
resulting from scattering of a single heavy quark on medium constituents
(``particle-hole" excitations), is known to impact the large-$\tau$
behavior of the in-medium correlators appreciably. Thus far, these
contributions have been estimated in quasiparticle approximation 
neglecting finite width 
effects~\cite{Aarts:2005hg,Umeda:2007hy,Mocsy:2007jz,Alberico:2007rg,Rapp:2008fv}.
In a treatment of heavy quarks consistent with the $T$-Matrix 
we evaluate the zero-mode contribution accounting for their full
spectral function, in particular finite-width effects. This automatically
yields predictions for the charm-quark number susceptibility, 
which can be compared to lattice data and enable a more reliable 
extraction of in-medium quasiparticle masses for heavy quarks.

The paper is organized as follows: In Sec.~\ref{sec_tmat} we give a 
short overview of the $T$-Matrix formalism employed in our calculations,
specifically discussing its extension to off-shell dynamics in
the charm-quark propagation. Section~\ref{sec_onia} is devoted to  
properties of cahrmonia in the QGP, i.e., their spectral 
functions including zero modes (Sec.~\ref{ssec_sf-zm}), euclidean 
correlator ratios in various mesonic channels (Sec.~\ref{ssec_corr})
and a comparison of numerical results using two different input 
potentials to lQCD data (Sec.~\ref{ssec_num-res}). 
In Sec.~\ref{sec_open} we first analyze the selfconsistently calculated 
charm-quark selfenergies (Sec.~\ref{ssec_sigma}) and then apply those to 
obtain the thermal relaxation rate and a schematic estimate of $\eta/s$ 
(Sec.~\ref{ssec_tau}), as well as the charm-quark number susceptibility 
(Sec.~\ref{ssec_chi}). We conclude in Sec.~\ref{sec_concl}.

%
\section{Off-shell $T$-Matrix at Finite Temperature}
\label{sec_tmat}
%
The formalism for calculating the $T$-Matrix for quark-quark and 
quark-antiquark scattering and bound states in the QGP, using two-body 
potentials estimated from heavy-quark free energies computed in lattice 
QCD (lQCD), has been developed in 
Refs.~\cite{Mannarelli:2005pz,Cabrera:2006wh,Riek:2010fk}. Here we recollect
the main elements while referring to Ref.~\cite{Riek:2010fk} for further 
details (e.g., a discussion of relativistic corrections and constraints from 
vacuum spectroscopy and the high-energy perturbative limit). Starting point 
is the Bethe-Salpeter Equation which after a 3-dimensional (3D) reduction 
and partial-wave expansion turns into a 1D integral equation for the 
scattering amplitude ($T$-Matrix), 
\begin{eqnarray}
T_{l,a}(E;q^\prime,q)=\mathcal{V}_{l,a}(q^\prime,q)
+\frac{2}{\pi}\int\limits^\infty_0 dk\,k^2\,\mathcal{V}_{l,a}(q^\prime,k)
\qquad
\nonumber\\
\times G_{12}(E;k)\,T_{l,a}(E;k,q)
 \ , 
\nonumber\\
\label{Tmat}
\end{eqnarray}
in a given color channel ($a$) and partial wave ($l$); the relative 
3-momentum moduli of the initial, final and intermediate 2-particle state 
are denoted by $q=|\vec{q}\,|$, $q^\prime=|\vec{q}^{\,\prime}\,|$ and 
$k=|\vec{k}\,|$, respectively (we restrict ourselves to vanishing total
3-momentum, $\vec P=0$, of the 2-particle system). 
The explicit form of the intermediate 2-particle 
propagator, $G_{12}$, depends on the 3D reduction 
scheme~\cite{Yaes:1971vw,Frohlich:1982yn,Blankenbecler:1965gx,Thompson:1970wt}  
(in slight deviation from 
Refs.~\cite{Mannarelli:2005pz,Cabrera:2006wh,Riek:2010fk} we here 
absorb the Pauli blocking factor, $(1-2f_F)$, into $G_{12}$). In the 
following, we focus on the Thompson scheme, since the Blankenbecler-Sugar 
scheme was found to generate some overbinding in the vacuum quarkonium 
spectrum~\cite{Riek:2010fk}. The $T$-matrix will be applied in both 
heavy-heavy and heavy-light quark channels for which a static 
(potential) approximation can be justified (i.e., the energy transfer 
is parameterically suppressed compared to the 3-momentum transfer, 
$\Delta q_0\sim (\Delta \vec q)^2/m_Q << \Delta q$, where
$\Delta \vec q = \vec q' -\vec q$).
For the two-body potential, $\mathcal{V}_{l,a}$, we follow our previous 
work~\cite{Riek:2010fk} using either the heavy-quark free ($F$) or internal 
($U$) energy computed in lQCD, implemented into a field-theoretical model 
for color-Coulomb and confining terms~\cite{Megias:2007pq} with 
relativistic corrections (e.g., the Breit 
interaction~\cite{Brown:2003km,Shuryak:2004tx,Brown:1952} to account 
for color-magnetic effects). To ensure the convergence of the Fourier 
transform from coordinate to momentum space we subtract the 
infinite-distance limit of the potential according to
\begin{equation}
V_a(r;T)=X_a(r,T)-X(r\to\infty,T) \ , \ \ X=F \ {\rm or} \ U \ ,
\label{V}
\end{equation}
and interpret $X_\infty(T)/2 \equiv X(r\to\infty,T)/2$ as a 
temperature-dependent correction to the bare HQ mass (real part of 
selfenergy) induced by the condensate associated with
the confining force term. The use of either $U$ or $F$ as potential is 
believed to bracket the uncertainties in this identification. 

\subsection{2-Particle Propagator}
\label{ssec_G12}
In all previous applications of potential models in the QGP the quarks
have been treated as quasiparticles with either vanishing or 
constant~\cite{Cabrera:2006wh,Riek:2010fk} width. However, as pointed out 
in Ref.~\cite{Riek:2010fk}, this approximation implies that quarkonium 
spectral functions, $\rho_\alpha(E)$ ($\alpha$: quantum numbers of 
the composite mesonic (or diquark) state), do not possess the proper 
low-energy limit, $\rho_\alpha(E\rightarrow0) \propto E$. While this 
is not expected to significantly impact the mass and binding energy of 
the bound states (for which the total energy is of order twice the
HQ mass), the euclidean correlators calculated below do actually involve
an integration over $\rho_\alpha(E)$ starting from $E=0$ with a thermal
weight which diverges for $E\rightarrow0$. In previous studies this 
problem has been evaded by introducing a low-energy cutoff on 
the spectral function (sufficiently below the lowest bound state as to 
not affect the correlator). However, recent studies of quark spectral 
functions~\cite{Beraudo:2009zz} find considerable low-energy strength in 
quarkonium spectral functions, e.g., due to particle-hole like structures 
in the HQ propagator. Thus a more elaborate treatment in the $T$-Matrix 
equation, properly accounting for off-shell dynamics, is in order. 

The general off-shell expression of the uncorrelated 2-fermion propagator 
at finite temperature, $G_{12}$, figuring into the $T$-matrix equation,
can be derived, e.g., within the Matsubara formalism. One has 
\begin{equation}
G_{12}(\Omega_\lambda,k) = T \sum_{\nu} G_1(z_\nu,\vec k) 
G_2(\Omega_\lambda-z_\nu,-\vec k)  
\label{G12-mats}
\end{equation}
where 
\begin{equation}
G_{i}(\omega,\vec k)=\frac{1}{\omega^2 - k^2 - m_i^2 - 2m_i\Sigma_i(\omega,k)} 
\end{equation}
($i=1,2$) denotes the (scalar part of) the 1-particle propagator, and
$z_\nu=i (2\nu+1)\pi T$ ($\Omega_\lambda$) are fermionic (bosonic) 
Matsubara frequencies. Using the spectral representations of the in-medium 
(retarded) single-particle propagators,
\begin{equation}
G_{i}(\omega,\vec k) = \int \frac{d\omega'}{2\pi} 
\frac{\rho_i(\omega',k)}{\omega-\omega'} \ , \quad \rho_i\equiv-2~{\rm Im} G_i \ ,
\end{equation} 
the Matsubara summation in Eq.~(\ref{G12-mats}) can be performed explicitly. 
After analytic continuation to the real axis ($\Omega_\lambda\to E$), the 
positive-energy contributions to the 2-particle propagator take the form
\begin{eqnarray}
G_{12}(E,k)&=&\int\frac{d\omega}{\pi}\int\frac{d\omega^\prime}{\pi}
\rho^{+}_1(\omega,k) \ \rho^{+}_2(\omega^\prime,k)
\nonumber\\
&& \ \times 
\,m_1\,m_2\,
\frac{1-f_F(\omega)-f_F(\omega^\prime)}{E-\omega-\omega^\prime+i\,\epsilon}
\label{G12}
\end{eqnarray}
where
\begin{eqnarray}
\rho^{+}_i(\omega,k)&=&\frac{-1}{\omega_{i}(k)}
{\rm Im}\frac{1}{\omega-\omega_{i}(k)-\Sigma_{i}(\omega,k)}
\nonumber\\ 
\omega_{i}(k)&=&\sqrt{k^2+m_{i}^2} \ 
\label{rho+}
\end{eqnarray}
denotes the positive-energy part of the quark spectral functions.
The factor $m_1 m_2$ in Eq.~(\ref{G12}) is specific to the Thompson 
reduction scheme, rendering the appropriate expression for $G_{12}$
in the limit of on-shell quarks~\cite{Thompson:1970wt}. 
For the in-medium HQ mass, as mentioned above (recall Eq.~(\ref{V})), 
the infinite-distance value of the HQ potential is identified with
a ``mean-field" contribution, \ie, as a real part of a selfenergy,  
\begin{equation}
m_Q=m_Q^0+\Sigma_Q^{\rm MF}(T) \ ,\ 
\Sigma_Q^{\rm MF}(T)\equiv X_\infty(T)/2 \ .
\label{m_eff}
\end{equation}
The additional selfenergy contribution figuring into the quark propagator
in Eq.~(\ref{rho+}) is generated from interactions with thermal light 
quarks and antiquarks and is therefore distinct from  $X_\infty(T)$.
It will be computed from the heavy-light quark $T$-matrix, as discussed
below. For calculating the latter, we need to specify light-quark
masses which we model by a thermal mass proportional to $gT$ with a 
coefficient to approximately match lQCD computations of the energy 
density of the QGP in quasiparticle approximation. In addition, we
supplement a moderate imaginary part for the light-quark selfenergy,
Im~$\Sigma_q=-0.05$~GeV, which is in the range of values obtained in 
Ref.~\cite{Mannarelli:2005pz} in a similar approach for light quarks, 
and comparable to what we obtain for heavy quarks (our results for 
quarkonium spectral function and HQ
transport are largely insensitive to the light-quark width).    

The evaluation of the Matsubara sum in Eq.~(\ref{G12-mats}) also
generates a contribution that solely arises from thermal
excitations (see, e.g., Ref.~\cite{Rapp:1995fv}). It involves one 
positive- and one negative-energy part of the quark spectral
functions which is why we referred to it as a ``particle-hole"
contribution above. For mesonic channels of the $T$-matrix, $12=Q\bar Q$,
its imaginary part can be cast into the form 
\begin{eqnarray}
{\rm Im}~G_{12}^{\rm ph}(E,k)= - \int \frac{d\omega}{2\pi} 
\rho_Q^+(\omega,k) \rho_Q^+(E+\omega,k) 
\nonumber\\
\times \left[ f^Q(\omega,k) - f^Q(E+\omega,k) \right] \ .
\label{G12-zm}
\end{eqnarray} 
Here, the negative-energy part of the $\bar Q$ spectral function
has been turned into the positive-energy part of the $Q$ spectral
function, i.e., the $\bar Q$ line in the original $Q\bar Q$ propagator
has been ``turned around". Together with the Fermi distributions, the 
interpretation of this contribution to $G_{12}$ becomes apparent: an 
incoming $Q$, pre-existing in the heat bath according to $f^Q(\omega,k)$,
scatters into a $Q$ with energy $E+\omega$, Pauli-blocked according
to $f^Q(E+\omega,k)$. $G_{12}^{\rm ph}$ is, in fact, precisely the 
zero-mode contribution discussed in the context of quarkonium 
correlators~\cite{Aarts:2005hg,Mocsy:2007jz,Umeda:2007hy,Alberico:2007rg}.
We return to its evaluation in Sec.~\ref{ssec_sf-zm} below.  

\subsection{Single-Quark Selfenergy}
\label{ssec_Sigma}
The HQ selfenergy, $\Sigma_{Q}(\omega,k)$, due to interactions with light 
anti-/quarks in the heat bath is calculated from the heavy-light ($Qq$) 
$T$-Matrix by closing its light-quark line with a quark propagator 
weighted with a Fermi distribution function. Using the Matsubara 
formalism one can express its imaginary part 
as~\cite{Rapp:1995fv,Mannarelli:2005pz}
\begin{eqnarray}
{\rm Im}~\Sigma_Q(\omega,k)&=&\frac{d_{SI}}{6k}\int\frac{p~dp}{(2\pi)^2}
\int\limits_{E_{\rm min}}^{E_{\rm max}} E~dE
\nonumber\\
&& \ \times {\rm Im}M_{qQ}(E,\omega,\omega_q(p),k,p)
\nonumber\\
&& \ \times\left[f_F(\omega_q(p))+f_B(\omega+\omega_q(p))\right]
\label{SigmaQ}
\end{eqnarray}
with
\begin{eqnarray}
&&M_{qQ}(E,\omega,\omega',k,p)=
\frac{m_{q}\,m_{Q}}{\omega_{q}(q_{cm})\,\omega_{Q}(q_{cm})}
\nonumber\\
&&\quad\times\,4\pi\sum_{a=1,8}d_{a}[T_{a,0}(E,q_{cm})+3T_{a,1}(E,q_{cm})]
\end{eqnarray}
where
\begin{eqnarray}
q_{cm}^2(E,k_{(4)},p_{(4)})&=&
\frac{(E^2-k_{(4)}^2-p_{(4)}^2)^2-4k_{(4)}^2\,p_{(4)}^2}{4E^2}
\nonumber\\
k_{(4)}^2&=&\omega^2-k^2 \quad , \quad  p_{(4)}^2=\omega'^2-p^2
\nonumber\\
E_{\rm min}^2&=&(\omega+\omega^\prime)^2-(k+p)^2
\nonumber\\
E_{\rm max}^2&=&(\omega+\omega^\prime)^2-(k-p)^2 \ .
\end{eqnarray}
The spin-isospin factor $d_{SI}=4N_f$ counts the degeneracy of available 
meson (or diquark) states, e.g., with total spin-0 and -1 for $S$-wave 
heavy-light scattering~\cite{vanHees:2007me}. In the expression for the HQ 
selfenergy, Eq.~(\ref{SigmaQ}), the thermal light quarks are treated as 
zero-width quasi particles so that their spectral functions can be replaced 
by $\delta$ functions (but with thermal mass $\sim gT$)\footnote{This 
is mainly done for numerical reasons. We have checked that using an 
off-shell quark spectral function leads to very similar results.}.
We recall that the gluonic contributions are entirely attributed to the 
mean-field type condensate term (for HQ transport, we include heavy-light
interactions to leading order in perturbation theory which does not
generate an imaginary part in the scattering amplitude). The real part 
of the selfenergy is obtained by a dispersion relation which is a 
preferred procedure in a selfconsistent prescription since the 
normalization of the spectral functions can easily be guaranteed. Our 
framework is similar to the one utilized in Ref.~\cite{Mannarelli:2005pz}
in the light-quark sector. However, in there the calculations of the 
$T$-matrix were restricted to on-shell selfenergies while here we 
use the full off-shell selfenergy of the heavy quark which, in particular,
enables to establish the correct low-energy behavior of the mesonic 
spectral functions (in addition to refinements in the implementation 
of the potential as developed in Ref.~\cite{Riek:2010fk} the non-potential 
corrections are expected to be significantly larger for light-quark 
interactions). 

Let us finally comment on the issue of imaginary parts in the
potential and selfenergy. Using effective field theories at finite 
temperature it has been found that the two-body potential operator
figuring into a Schr\"odinger equation acquires an imaginary
part~\cite{Beraudo:2007ky,Brambilla:2008cx,Laine:2006ns,Rothkopf:2009pk}.
Diagrammatically, this implies that the potential possesses on-shell cuts 
corresponding to dissociation processes of the composite (bound) state. 
Let us inspect two commonly discussed (``leading-order") processes for the 
case of charmonium bound states. For large binding energy, $E_B\ge T$, 
the dominant process is gluon dissociation, $g+J/\psi \to c+\bar c$, 
first analyzed by Bhanot and Peskin~\cite{Bhanot:1979}. In the 
language of effective field theory, this corresponds to the so-called 
color-singlet to -octet transition mechanism (up to final-state 
interactions in the octet channel, which, however, are suppressed by 
$1/N_c^2$ and thus numerically negligible).
In our $T$-matrix formalism, the inclusion of this process would require
a coupled-channel treatment, with a $c\bar c$-gluon intermediate state, 
whose cut exactly produces the corresponding decay channel. Such a 
calculation is beyond the scope of the present work. For small binding 
energies, $E_B < T$ (as relevant for excited states or the case of
reduced in-medium $J/\psi$ binding energies close to its dissolution 
temperature), gluo-dissociation is phase-space suppressed and thus 
superseded by the Landau-damping phenomenon in the (spacelike) 
one-gluon exchange of the potential. This is precisely the imaginary 
part of the two-body potential discussed
in Refs.~\cite{Beraudo:2007ky,Laine:2006ns,Rothkopf:2009pk}. Physically, 
it corresponds to ``quasifree" dissociation~\cite{Grandchamp:2001pf}, 
$p+J/\psi \to p + c + \bar{c}$, induced by thermal partons 
$p=g,q,\bar{q}$ (which generate the imaginary part of the one-loop 
correction to the exchanged gluon). In the limit of small binding (or 
large charmonium size), the incoming thermal partons with energy 
$\sim$$T$ do not sense the size of the bound state and thus the 
scattering effectively happens on the level of an individual charm
(or anti-charm) quark. Thus, in the $T$-matrix formalism this process
is encoded in the selfenergy of a single (anti-) charm quark, which is
included in our calculations. For large binding, quasifree dissociation
ceases since the thermal parton does not resolve the substructure of
the color neutral bound state~\cite{Beraudo:2007ky,Laine:2006ns}. 
On the other hand, when formulated as a potential contribution, 
the large-distance limit of its imaginary part coincides with twice 
the imaginary part of the charm-quark selfenergy~\cite{Beraudo:2007ky}. 
Therefore our evaluation of the correlators within the $T$-Matrix 
approach using a real potential is well in line with the use of a 
complex potential in a Schr\"odinger treatment.
In the $T$-matrix, imaginary parts are generated via unitarization of
intermediate (on-shell) states including single-particle selfenergy 
contributions.

\section{Quarkonia}
\label{sec_onia}
In this section we apply the selfconsistent $T$-matrix to charmonia 
with special consideration of the zero-mode contribution 
(Sec.~\ref{ssec_sf-zm}) to the Euclidean correlators in different
mesonic channels, $\alpha=S,PS,V,AV$, i.e., scalar, pseudoscalar,
vector and axialvector, respectively (Sec.~\ref{ssec_corr}),
followed by a discussion of numerical results using either $U$ or $F$ 
as underlying two-body potential (Sec.~\ref{ssec_num-res}).

\subsection{Spectral Functions and Zero Modes}
\label{ssec_sf-zm}
With the $c\bar c$ $T$-Matrix from Eq.~(\ref{Tmat}) we proceed to 
determine the charmonium spectral function including both bound and 
scattering states as in Ref.~\cite{Riek:2010fk}. The $T$-matrix
signifies the rescattering contribution to the correlation
function which is schematically given by~\cite{Cabrera:2006wh}
\begin{equation}
G = G^0  + G^0 T G^0  
\end{equation}
where $G^0$ denotes the free 2-particle loop. This can be 
be represented diagrammatically as
\begin{eqnarray}
G=\parbox{1.2cm}{\includegraphics[scale=0.5]
{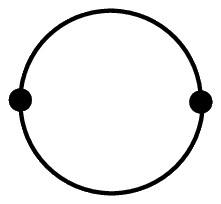}}+\parbox{3cm}
{\includegraphics[scale=0.5]{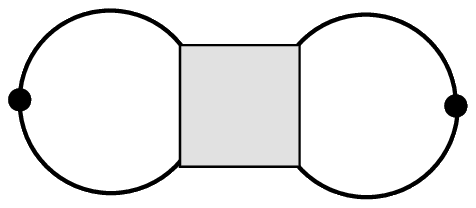}} \, 
\end{eqnarray}
where compared to Eq.~(\ref{G12}) the loop now also includes an 
integration over relative momentum, $\vec k$, as well as vertices 
(denoted by dots) specifying the quantum-number channel $\alpha$. The 
spectral function is then defined as usual by $\rho=-2{\rm Im}G$.  

Quantitative comparisons to euclidean correlator ratios as ``measured" 
in lQCD~\cite{Jakovac:2006sf,Aarts:2007pk} require the inclusion of 
zero-mode 
contributions~\cite{Umeda:2007hy,Aarts:2005hg} which turn out to be 
different for different meson channels  $\alpha$. This, in particular, 
lifts the spin degeneracy within $S$-wave ($PS-V$) and $P$-wave 
($S-AV$) states of the $T$-matrix. This is not surprising since HQ 
symmetry is not expected to be valid for zero modes. A relativistic 
evaluation of the vertices figuring into 
$\rho_\alpha^{\rm zm}\equiv -2{\rm Im}~G_\alpha^{\rm ph}$ is therefore 
in order. To be consistent with the treatment of the $T$-Matrix we 
evaluate this contribution beyond the zero-width quasiparticle
approximation (as applied in the literature to date) by taking into 
account the finite width of the quark. As in previous literature, we 
focus on the leading part of the zero-mode contribution which does not
involve a two-body potential contribution\footnote{A non-perturbative 
treatment of the particle-hole interaction would require a relativistic 
treatment solving the Bethe-Salpeter-Equation which goes beyond the 
potential approximation adopted here.}. Augmenting Eq.~(\ref{G12-zm})
with relativistic vertices one obtains~\cite{Aarts:2005hg}
\begin{eqnarray}
\rho^{\rm zm}_{\alpha}(P)&=&{2N_{c}}
\int\frac{d^{4}k}{(2\pi)^{4}}\,{\rm Tr}
\left[(\kslash+m_c)\,\Gamma\,(\rslash +m_c)\,\Gamma^{\dagger}
\right]
\nonumber\\
&&\times\,\rho_{c}^+(k)\,\rho_{c}^+(r)\,\left[f^{F}(k^{0})-f^{F}(r^{0})\right]
\nonumber\\
&=&2N_{c}\int\frac{d^{4}k}{(2\pi)^{4}}\,4\,
F_{\Gamma}(k,r,m_c)
\nonumber\\
&&\times\,\rho_{c}^+(k)\,\rho_{c}^+(r)\,\left[f^{F}(k^{0})-f^{F}(r^{0})\right]
\label{rho-zm}
\end{eqnarray}
where we introduced the notation
\begin{equation}
r=P+k \quad  , \quad
\Gamma\in\{1,\gamma_5\,,\gamma_0,\gamma_i,\gamma_5\gamma_i\}
\end{equation}
with $r$, $k$ and $P=(E,\vec P)$ denoting 4-vectors for the purposes 
of the above
two and the following equation only. The kinematic factors arising from
the different Dirac structures take the form  
\begin{eqnarray}
F_{1}(k,r,m)&=&k\cdot r+m_c^2
\nonumber\\
F_{\gamma_5}(k,r,m)&=&k\cdot r-m_c^2
\nonumber\\
F_{\gamma_0}(k,r,m)&=&k_0\cdot r_0+\vec{k}\cdot\vec{r}+m_c^2
\nonumber\\
F_{\gamma_i}(k,r,m)&=&3\,r^0k^0-\vec{r}\cdot\vec{k}-3m_c^2 
\nonumber\\
F_{\gamma_5\gamma_i}(k,r,m)&=&3\,r^0k^0-\vec{r}\cdot\vec{k}+3m_c^2 \ ;
\label{Fzm}
\end{eqnarray}
a summation over spatial indices $i$ is implied for the vector and 
axialvector case. In the previously adopted zero-width approximation 
the zero-mode spectral function can be simplified to~\cite{Aarts:2005hg}
\begin{equation}
\rho_{\alpha}^{\rm zm}(E,\vec{P}\rightarrow 0)
=2\pi E\,\delta(E)\chi_{\alpha}(T)
\end{equation}
where 
\begin{equation}
\chi_{\alpha}(T)=-\,2N_{c}\int \frac{d^{3}\vec{k}}{(2\pi)^3}
\left(c_1+c_2\frac{\vec{k}^2}{\omega_{c}(\vec{k})^2}\right)
\frac{\partial f^{c}(\omega_{c}(\vec{k}))}{\partial\omega_{c}(\vec{k})}
\label{chi_zero_width}
\end{equation}
denotes a generalized susceptibility with coefficients corresponding
to different quantum number channels according to
\begin{center}
\begin{tabular}{lcllclcll}
$\Gamma=1$ & $\,\Rightarrow\,$ & $c_{1,2}=2,-2$ & \phantom{xxx} & 
$\Gamma={\gamma_5}$ & $\,\Rightarrow\,$ & $c_{1,2}=0,0$\\
$\Gamma={\gamma_0}$ & $\,\Rightarrow\,$ & $c_{1,2}=2,0$ &  & 
$\Gamma={\gamma_i}$ & $\,\Rightarrow\,$ & $c_{1,2}=0,2$\\
$\Gamma={\gamma_5\gamma_i}$ & $\,\Rightarrow\,$ & $c_{1,2}=6,-4$ \ .
\end{tabular}
\end{center}
In our numerical calculations reported below we evaluate 
$\rho_\alpha^{\rm zm}$ directly from Eq.~(\ref{rho-zm}) using the
positive-energy charm-quark spectral function, Eq.~(\ref{rho+}), with
the selfconsistently determined off-shell selfenergy. This puts 
the treatment of the zero-mode contribution on the same level as 
the $c\bar c$ scattering and bound-state part using
the 2-particle propagator $G_{12}$ in Eq.~(\ref{G12}).

\subsection{Euclidean Correlator Ratios}
\label{ssec_corr}
The transformation of the spectral function to the euclidean correlator
is given by
\begin{eqnarray}
G_\alpha(\tau,T)&=&\int \frac{dE}{2\pi} \,
\left[\rho_{\alpha}(E,T)+\rho^{\rm zm}_{\alpha}(E,T)\right]
\,\mathcal{K}(\tau,E,T)
\nonumber\\ 
\mathcal{K}(\tau,E,T)&=&
\frac{\cosh\left[E\left(\tau-1/2T\right)\right]}
{\sinh\left[E/2T\right]} \ . 
\label{eucl-corr}
\end{eqnarray}
As before, we focus an vanishing total 3-momentum, $\vec P=0$, of 
the $Q\bar Q$. 
The temperature kernel $\mathcal{K}$ imprints an exponential $\tau$
dependence on the correlator. To better exhibit the in-medium 
modifications of $G_\alpha(\tau,T)$ it is therefore common to 
analyze the correlator ratio
\begin{equation}
R_G^\alpha(\tau,T)=\frac{G_\alpha(\tau,T)}{G^{\rm rec}_\alpha(\tau,T)}
\end{equation}
where the denominator is the so-called reconstructed correlator
which is evaluated with the kernel $\mathcal{K}$ at temperature $T$
but with a ``reference" spectral function typically taken as the
vacuum one (or at low temperature $T^*$ where no significant medium
modifications are expected). 

\subsection{Numerical Results}
\label{ssec_num-res}
\begin{figure*}[!t]
\includegraphics[scale=0.7]{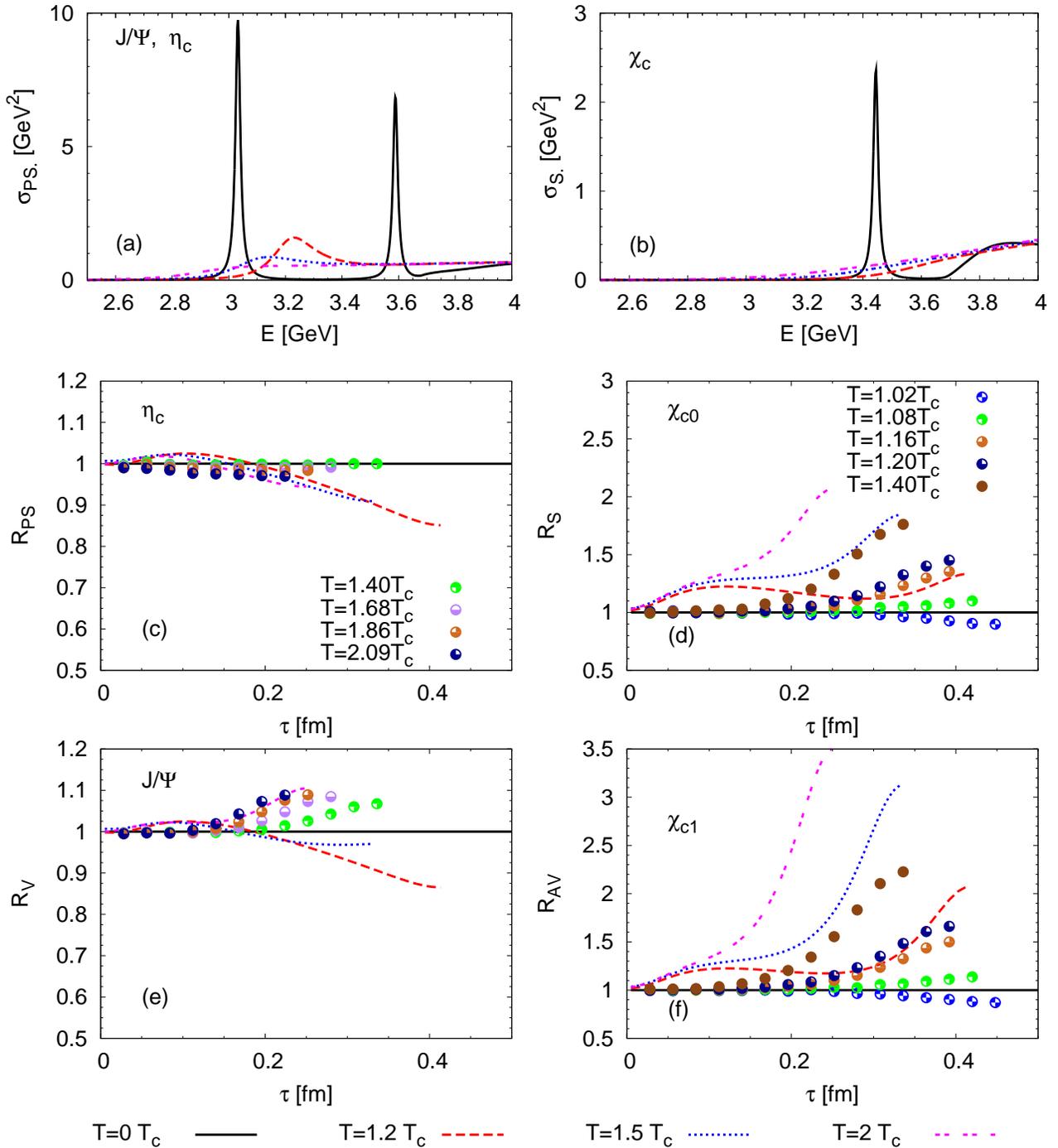}
\caption{(Color online) Charmonium spectral functions in the pseudoscalar
(upper left panel) and scalar (upper right) channel for different
temperatures using the internal energy, $U$, as $Q\bar Q$ potential. 
In the middle and lower rows we show the corresponding euclidean 
correlator ratios in the pseudoscalar (middle left), scalar (middle 
right), vector (lower left) and axialvector channels (lower right), 
where the latter three include zero-mode contribution. The comparison 
to the lattice data (indicated by symbols)~\cite{Aarts:2007pk} is made 
in absolute units of $\tau$ (limited to $1/2T$ at each temperature), 
with $T_c\simeq210$\,MeV for the lattice data and $T_c\simeq196$\,MeV 
underlying the potential for the $T$-matrix calculations (e.g., 
1.4~$T_c^{\rm lat}\simeq1.5~T_c^{\rm Tmat}$).}
\label{fig_Spec-U}
\end{figure*}
\begin{figure*}[!t]
\includegraphics[scale=0.7]{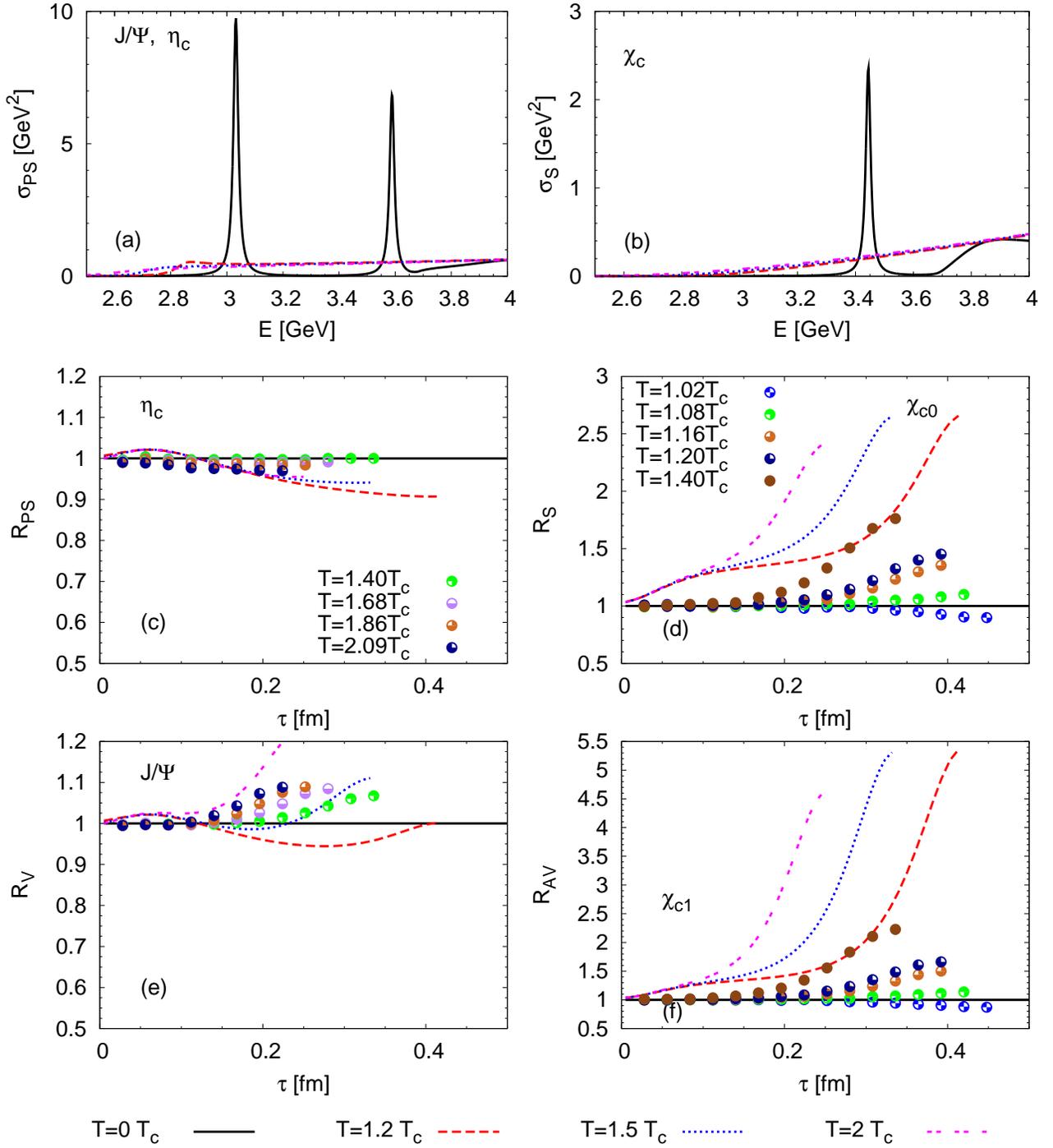}
\caption{(Color online) Same as Fig.~\ref{fig_Spec-U}
but using the free energy, $F$, as $Q\bar Q$ potential.}
\label{fig_Spec-F}
\end{figure*}
Before turning to the numerical results for the finite-$T$ spectral
functions and correlator ratios let us briefly summarize our input
quantities, largely as given in Ref.~\cite{Riek:2010fk} (where more 
details can be found, including extensive analysis of the associated 
uncertainties). Our potentials ($F$ or $U$) are based on fits to the 
lattice results for (2+1)-flavor QCD from
Refs.~\cite{Kaczmarek:2007pb,Cheng:2007jq,Kaczmarek-private} 
(``potential 1" in Ref.~\cite{Riek:2010fk}). In vacuum our off-shell
calculations reproduce the results of Ref.~\cite{Riek:2010fk} since in 
the limit of vanishing $c$-quark width the propagator $G_{12}$ reduces 
to the standard Thompson form. The bare charm-quark mass is set to 
$m_c^0=1.264$~GeV, which, together with a vacuum selfenergy of 
$X_\infty^{\rm vac}/2=0.6$\,GeV, gives a fair description of the 
spin-average masses of $J/\psi$-$\eta_c$, $\psi'$ and $\chi_c$ states.

The full off-shell treatment with more realistic $c$-quark propagators
in the present work leads to moderate but significant changes for the
in-medium results. In addition, the zero-mode contributions in the $V$,  
$S$ and $AV$ channels (recall Eq.~(\ref{rho-zm})) have a marked impact 
on the corresponding correlator ratios. For a comparison to lQCD 
correlator ratios we choose the results of the $N_f=2$ computations 
from Ref.~\cite{Aarts:2007pk}, which correspond more closely to our 
input potentials than quenched calculations. The critical temperature 
in the simulations of the lQCD correlator ratios~\cite{Aarts:2007pk} 
is $T_c\simeq210$\,MeV, which is not far from the one underlying our 
potential~\cite{Kaczmarek:2007pb}, $T_c\simeq196$\,MeV. Therefore,
rather than normalizing to the different $T_c$'s, we compare our
results for the correlator ratios with the lQCD data in absolute
units of euclidean time, $\tau$. Each correlator ratio is plotted
up to the midpoint, $\tau=1/2T$, of the total $\tau$ range (the
correlators are symmetric about this point). Thus, comparable
temperatures are easily identified by the same $\tau$ range in the
plotted ratio (if one were to normalize all results to $T_c$, lQCD 
data at given absolute temperature would be compared to $T$-matrix
calculations at slightly lower $T$).
 
The in-medium results using the internal energy, $U$, as potential are
summarized in Fig.~\ref{fig_Spec-U} in terms of the $S$- and $P$-wave
spectral functions in the upper panels (degenerate for $\eta_c$-$J/\psi$ 
and for $\chi_c$ states) and the correlator ratios in the middle
and lower panels (where the degeneracies are lifted by the zero modes). 
Since the (magnitude of the) imaginary part of the selfconsistently 
calculated $c$-quark selfenergy turns out to be around 0.050-0.100~GeV 
for low-momentum on-shell charm quarks (cf.~Fig.~\ref{fig_sigma-c} 
below), the closest comparison to our earlier quasiparticle results 
is for the case of a constant (energy and 
3-momentum independent) imaginary part of  Im\,$\Sigma_Q=-0.05$~GeV 
(Fig.~14 in Ref.~\cite{Riek:2010fk}). In the $S$-wave spectral function
we find slightly more attraction for the ground-state peak for
temperatures below 1.5\,$T_c$ (for higher $T$ it dissolves, as in 
Ref.~\cite{Riek:2010fk}). Despite the larger on-shell 
width in our present treatment, which is up to twice as large as in our 
previous quasiparticle calculations, the width of the $J/\Psi$-$\eta_c$ 
peaks is very similar. This is a direct consequence of the energy and 
momentum dependence of the selfenergy which decreases considerably 
off-shell (cf.~Fig.~\ref{fig_sigma-c} below) and thus reduces the 
``operative" width of $c$-quarks in $c\bar c$ bound states. It becomes 
apparent in the $\eta_c$ correlator ratio (where no zero-mode is active) 
which at the lowest considered temperature (1.2\,$T_c$) drops to about 0.85
compared to 0.9 in Ref.~\cite{Riek:2010fk}, even though the peak position 
of the bound state is shifted to slightly lower energies in the present 
calculation (which tends to increase the large-$\tau$ correlator 
ratio). This finding shows that a proper off-shell treatment is warranted 
to correctly account for the low-energy strength in the spectral function
which has significant impact on the correlator ratio at large $\tau$. 
We are not too concerned that the drop to 0.85 is noticeably larger than 
in the lQCD data since we have neglected several effects which will 
contribute further low-energy strength to the spectral function, e.g.,  
imaginary parts in the $c$-quark selfenergy from scattering off thermal 
gluons, or coupled channels in the $c\bar c$ $T$-matrix such as $D\bar D$ 
and $c\bar c g$ (inducing singlet-to-octet transitions). These are expected 
to be especially relevant close to $T_c$ ($D$-meson states will form close 
to $T_c$ and gluo-dissociation is efficient for large charmonium binding,
$E_B\ge T$). On the other hand, we note that the $PS$ correlator ratio
is remarkably independent of temperature and closer to one for intermediate 
and small $\tau$ than in Ref.~\cite{Riek:2010fk} (e.g., no more than 
$\sim$2\% above one), which improves the agreement with lQCD.
In the $V$, $S$ and $AV$ channels the zero-mode contributions lead
to a marked enhancement of the correlator ratios at large $\tau$,
especially for the $P$-wave channels where the $c_1$ coefficient is
non-zero (for the vector channel, we sum over the spatial components
only, corresponding to $\Gamma=\gamma_i$ in Eq.~(\ref{rho-zm}). Compared 
to a zero-width treatment of the zero-mode contribution, corresponding 
to Eq.~(\ref{chi_zero_width}), the inclusion of a finite quark width 
increases the correlator ratios at large $\tau$ by ca.~0.1-0.15.
Overall, the agreement of the the calculated correlator ratios 
with $N_f=2$ lQCD data~\cite{Aarts:2007pk} included in the middle and 
lower panels of Fig.~\ref{fig_Spec-U} is fair. The largest discrepancies 
of ca.~$\sim$30\% at both intermediate and large $\tau$ occur in the 
$AV$ ($\chi_{c1}$) channel where the zero-mode contribution is the 
strongest. The extracted melting temperatures of the charmonium bound 
states are close to our earlier determination with finite but constant 
width~\cite{Riek:2010fk}, \ie, ca.~1.5\,$T_c$ for the $J/\Psi$ and 
below 1.2\,$T_c$ for all other states ($\Psi^\prime$, $\chi_c$).

The results obtained using the free energy, $F$, as potential are compiled 
in Fig.~\ref{fig_Spec-F}. As in previous 
work~\cite{Riek:2010fk,Mocsy:2005qw,Mocsy:2007jz}, we find a 
much stronger suppression of the bound states compared to using $U$, with 
a melting temperature for the $S$-wave ground state of 1.2\,$T_c$ or even 
lower, whereas the $P$-wave spectral function is already structureless 
at this temperature. Nevertheless, the correlator ratio in the $\eta_c$ 
channel (no zero mode) is surprisingly $T$-stable and close to one: the 
loss of low-energy strength due to the dissolved bound state is compensated
by a reduced $c\bar c$ threshold in connection with a nonperturbative
rescattering strength in the threshold region generated through the
$T$-matrix~\cite{Cabrera:2006wh}. Indeed, the in-medium charm-quark mass 
(correction) following from Eq.~(\ref{m_eff}) for $F_\infty$ is 
significantly smaller than for $U_\infty$. This, however, generates a
markedly enhanced zero-mode contribution to the correlator ratios in the
$V$, $S$ and $AV$ channels, compared to the case with $U$ as potential 
(the effect of the selfconsistently calculated finite quark width in the 
zero modes is a ca.~10\% enhancement at large $\tau$).    
In particular in the $P$-wave channels, the large-$\tau$ is quite a bit 
off the $N_f$=2 lQCD data. Apparently a potential closer to the internal 
energy is favored (with larger $m_c^*$ but stronger $c\bar c$ binding).

\section{Open-Charm Transport}
\label{sec_open}
We now turn to examining the numerical results for the single charm-quark
properties in the QGP, specifically its selfenergy (mass correction and
scattering rate; Sec.~\ref{ssec_sigma}), thermal relaxation rate 
(Sec.~\ref{ssec_tau}) and number susceptibility (Sec.~\ref{ssec_chi}). 
We recall that these are selfconsistently computed via numerical
iteration with the $T$-matrix in all heavy-light quark channels,
Eq.~(\ref{Tmat}).

\subsection{Selfenergy}
\label{ssec_sigma}
\begin{figure*}[t]
\includegraphics[scale=0.7]{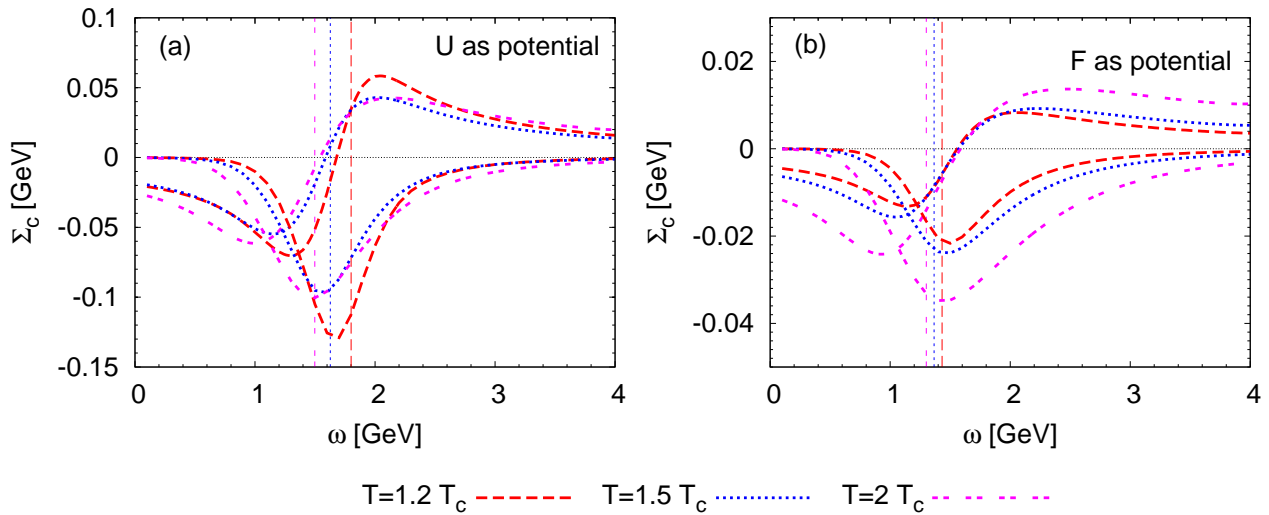}
\caption{(Color online) Real and imaginary parts of 
the charm-quark selfenergy at vanishing 3-momentum in a QGP at different 
temperatures for the scenarios where the internal (left panel) or free 
energy (right panel) is employed as potential. The imaginary parts are 
negative definite while the real parts change sign close to the vertical 
lines which indicate the effective charm-quark mass (on-shell energy) at 
the respective temperature.}
\label{fig_sigma-c}
\end{figure*}
The real and imaginary parts of the charm-quark selfenergy following 
from the $T$-Matrix are displayed in Fig.~\ref{fig_sigma-c} as a function 
of quark energy for vanishing 3-momentum and for three temperatures. The
general structure is that of a maximum around the on-shell energy,
$\omega\simeq m_c^*$, in the imaginary part, associated with a typical
zero crossing in the real part. The peak structure is caused by Feshbach
type resonances (or threshold enhancements) around the heavy-light quark 
threshold in the $T$-matrix~\cite{vanHees:2007me,Riek:2010fk}. These 
features are the reason that, upon (off-shell) integration over $\omega$ 
in the two-particle propagator ($T$-matrix), the contributions from the 
real part largely cancel while the width effects drop significantly
faster (and thus on average are smaller) than for a fixed (constant) 
quasiparticle particle width as employed in previous 
work~\cite{Cabrera:2006wh,vanHees:2007me,Riek:2010fk}.
 
More quantitatively, in the case of $U$ as potential (left panel in
Fig.~\ref{fig_sigma-c}), the on-shell width of the charm quark reaches 
a value of up to $\Gamma_c=-2\,{\rm Im}\Sigma_c\simeq250$\,MeV at the 
lowest considered temperature (1.2\,$T_c$), quite similar to what has been 
calculated in Ref.~\cite{vanHees:2007me} using different lQCD inputs for 
$U$. The magnitude decreases with increasing temperature (not as much
as in Ref.~\cite{vanHees:2007me}), which is quite remarkable given the
fact that the thermal densities of the thermal anti-/quarks decrease
appreciably (in the ``$T$-$\varrho$" approximation, the selfenergy would
be directly proportional to the light-quark density, 
$\Sigma_c\sim T_{cq} n_q$). It clearly reflects the weakening of the
two-body interaction as temperature increases, due to color-charge
screening in both the Coulomb and confining parts of the potential. 
In other words, the maximal interaction strength is realized at the
lowest temperature, just above $T_c$.  

For the $F$-potential the $c$-quark selfenergy is reduced by 
about a factor of 6 at the lowest temperature, which reduces to a factor 
of $\sim$2-3 at 2\,$T_c$. In this scenario, no Feshbach resonances form, 
but rescattering embodied in the $T$-matrix still produces a notable 
threshold enhancement which induces an on-shell width of ca.~40-70\,MeV.

\subsection{Thermal Relaxation Rate}
\label{ssec_tau}
%
\begin{figure*}[t]
\includegraphics[scale=0.7]{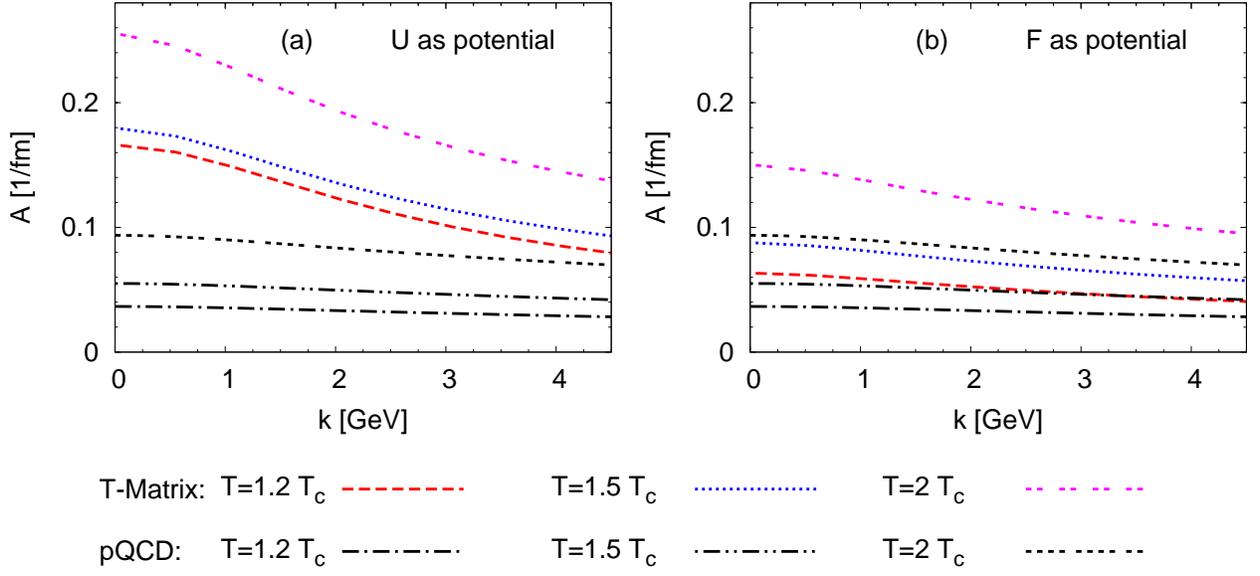}
\caption{(Color online) Charm-quark relaxation rate as a function of 
3-momentum for different temperatures using either the $T$-Matrix plus 
pQCD gluon scattering ~\cite{Riek:2010fk} (upper 3 lines in 
left panel) or a pQCD calculation for scattering off gluons and 
anti/quarks (lower 3 lines in left panel). The $T$-matrix is computed
using $U$ (left) or $F$ (right) as potential while pQCD contributions
are obtained with $\alpha_s$=0.4.}
\label{fig_A}
\end{figure*}
Next we turn to the calculation of the thermal relaxation rate of charm 
quarks in the QGP, employing a Fokker-Planck treatment following
Ref.~\cite{Svetitsky:1987gq}. The relation of the drag coefficient, $A(p)$, 
to the $T$-matrix has been elaborated in Ref.~\cite{vanHees:2007me}, 
including all color configurations in $c$-$q$ and $c$-$\bar q$ scattering
in $S$- and $P$-waves. For a more realistic evaluation of the total
coefficient, we add to the $T$-matrix contribution the effect of 
$c$-quark scattering off thermal gluons which we approximate with 
the leading order perturbative diagrams (using $\alpha_s$=0.4) including 
Debye screening masses in the exchange propagators ($t$-channel gluon 
exchange gives the dominant contribution). In Fig.~\ref{fig_A} the full 
results for $A(p)$
are compared to perturbative calculations in which scattering off both 
anti-/quarks and gluons is obtained from the LO diagrams. Compared
to the quasiparticle treatment in Ref.~\cite{Riek:2010fk}, the 
full off-shell treatment leads to a ca.~10\% increase of the drag 
coefficient at low 3-momenta, while the deviations are small 
at momenta of $k\ge2$\,GeV. This is well in line with the finding in 
Ref.~\cite{Riek:2010fk} that the dependence of the drag coefficient on
the 3D reduction scheme of the $T$-matrix equation (relating to its
off energy-shell behavior via different 2-particle propagators) is small.
Thus, we confirm as a robust feature that $c$-quark thermalization using 
the nonperturbative $T$-matrix is accelerated over pQCD calculations
by about a factor of 4 (2) when using $U$ ($F$) as potential. 

\begin{table}[t]
$\gamma_c$ [1/fm] with $U$-Potential\\
\begin{tabular}{l|c|c|c}
\hline         & Heavy-Light & Quarkonium & \\
 $T$ [$T_c$]   & $T$-Matrix & spectral function & ``$\eta/s$"\\
\hline 1.2 & 0.137 & 0.177 & 0.16-0.2\\
\hline 1.5 & 0.136 & 0.148 & 0.28-0.3\\
\hline 2.0 & 0.180 & 0.147 & 0.41-0.47\\
\end{tabular}
                                                                            
\vspace{0.5cm}
                                                                            
$\gamma_c$ [1/fm] with $F$-Potential\\
\begin{tabular}{l|c|c|c}
\hline               & Heavy-Light& Quarkonium & \\
       $T$ [$T_c$]   & $T$-Matrix & spectral function & ``$\eta/s$"\\
\hline 1.2 & 0.034 & 0.034 & 0.6\\
\hline 1.5 & 0.044 & 0.036 & 0.73-0.8\\
\hline 2.0 & 0.075 & 0.051 & 0.8-0.96\\
\end{tabular}
\caption{Comparison of charm-quark relaxation rates, $\gamma_c(k=0;T)$, 
as obtained from the heavy-light $T$-Matrix~\cite{Riek:2010fk} 
(2.\,column) and from a Kubo formula for the diffusion coefficient 
(3.\,column) using the low-energy limit of the heavy-quarkonium 
spectral function, Eq.~(\ref{gamma_c}), for the two different scenarios 
for potential (upper and lower table). 
The last column lists a schematic estimate of the 
viscosity to entropy-density ratio using the kinetic-theory relation,
Eq.~(\ref{eta-s}), including the contributions from scattering off
thermal gluons (the quoted ranges only reflect the variation in the
2 preceding columns; the true uncertainty is larger).}
\label{comp_gamma_c}
\end{table}
In addition to the direct calculation of the relaxation rate following 
from the collision term in the Boltzmann (or Fokker-Planck) equation
with the heavy-light $T$-matrix~\cite{vanHees:2007me,Riek:2010fk}, 
the off-shell framework set up in the present paper enables an 
alternative method,
namely from the zero-energy limit of the heavy-quarkonium spectral 
function~\cite{Petreczky:2005nh,CaronHuot:2009uh},
\begin{eqnarray}
\gamma_c=\frac{T}{m_c\,D_s}\qquad D_s=\frac{1}{\chi_{c}(T)}
\lim_{\omega\rightarrow 0}\frac{\rho_{\gamma_i}(\omega,0)}{2\omega} \ 
\label{gamma_c}
\end{eqnarray}
($\chi_c\equiv\chi_{00}$: charm-quark number susceptibility, 
$D_s$: spatial diffusion constant).
The values extracted from this method correspond to the zero-momentum
limit and are compared to the pertinent values from the collision
integral using the heavy-light $T$-matrix in Tab.~\ref{comp_gamma_c}
(for a consistent comparison the contribution from scattering off 
thermal gluons is not included). Note that the latter involves the
$Qq$ scattering amplitude squared, while the imaginary part of the 
quarkonium spectral function is basically determined by the imaginary 
part of the single-quark spectral function which, in turn, is obtained 
from the imaginary part of the heavy-light scattering amplitude, 
cf.~Eq.~(\ref{SigmaQ}). We find that the extraction from
the low-energy limit of the charmonium spectral function in the 
vector channel tends to give larger (smaller ) values at lower 
(higher) for both $U$ or $F$ as two-body potential.   
One of the uncertainties which is presumably reflected by these
deviations is the Fokker-Planck approximation when using the
heavy-light $T$-matrix to evaluate the diffusion coefficient. For 
example, in Ref.~\cite{vanHees:2004gq} it has been found that a 
$D$-meson resonance model in the QGP can lead to a violation of the 
Einstein relation by underestimating of $\gamma_c$ by up to 10\% at 
temperatures of $T\simeq300$\,MeV, while the agreement is 
closer at lower temperatures (less than 5\% for $T<250$\,MeV; 
note that the $D$-meson like correlations in the $T$-matrix dissolve
for temperatures above 1.5\,$T_c\simeq 300$\,MeV). Overall, the
discrepancies between the two methods are within $\sim$30\%, which, 
given the different schemes of evaluating this transport 
coefficient, is not too bad. The systematic trends are similar
when using $F$ or $U$ as potential, and the large difference
between these 2 scenarios is robust. 

\begin{figure*}[!t]
\includegraphics[scale=0.71]{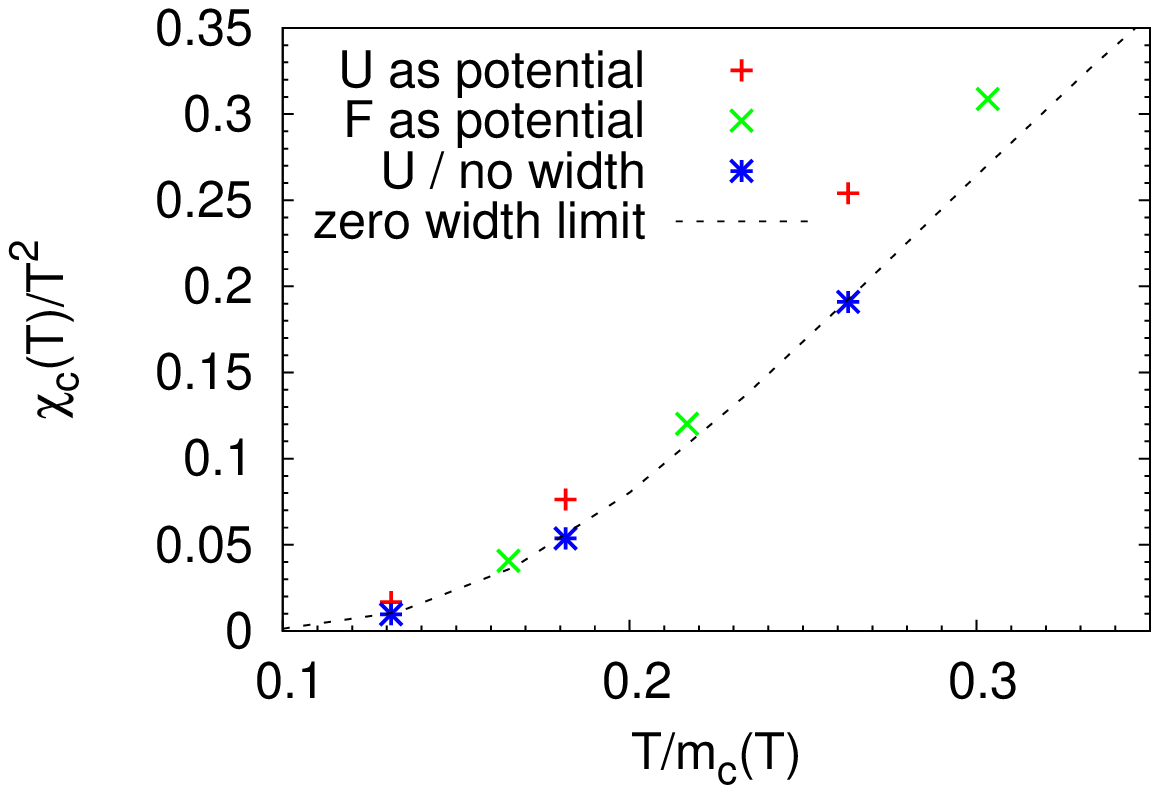}
\includegraphics[scale=0.69]{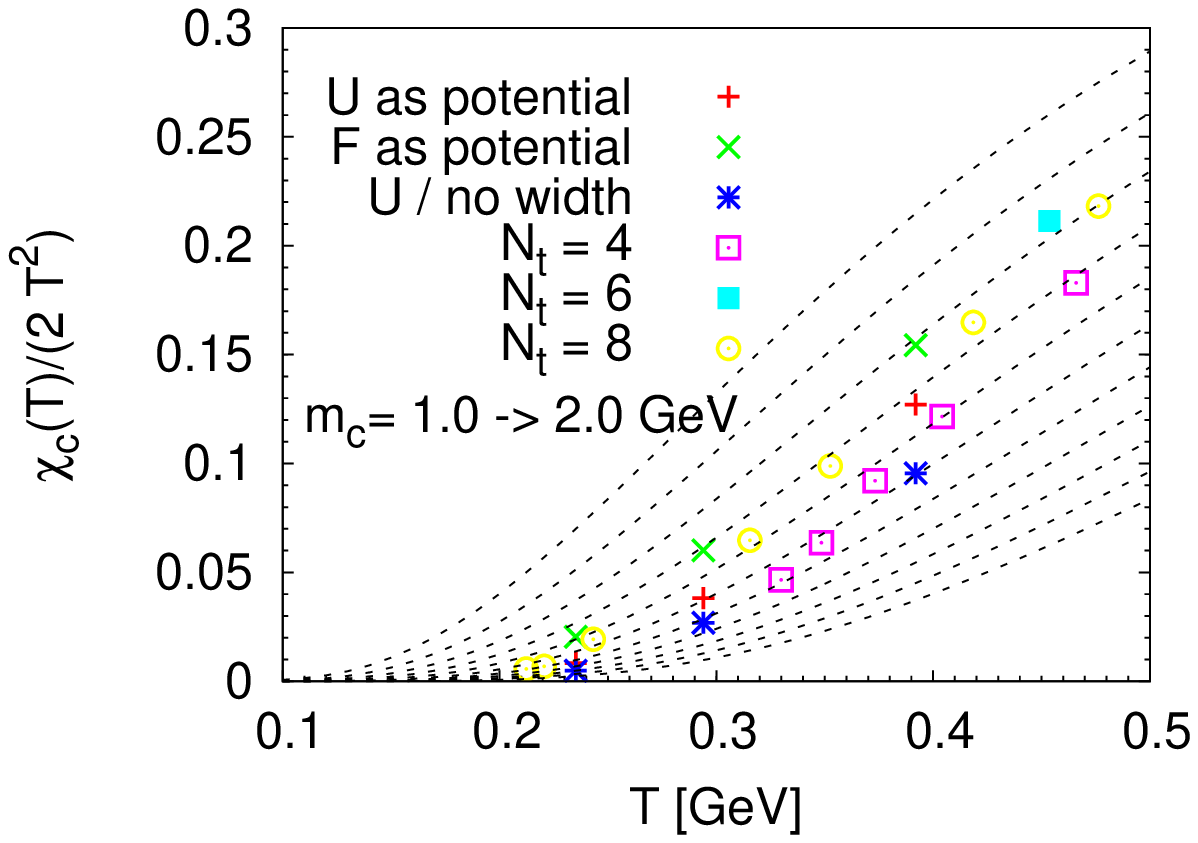}
\caption{(Color online) Charm-quark number susceptibility as a 
function of temperature normalized to the zero mass limit, 
Eq.~(\ref{chi_zero_mass}). Left panel: comparison of our full
off-shell calculations with finite width (using either $U$ or
$F$ as potential) to the zero width-limit using $U$ and the
quasiparticle expression, Eq.~(\ref{chi_zero_width}), with $T$-dependent
$c$-quark mass; the temperature on the $x$-axis has been 
rescaled by the respective in-medium masses in each scenario.
Right panel: comparison of our results to lattice computations
in 2+1-flavor QCD with different numbers, $N_t$, of lattice points
in temporal direction; the dashed lines indicate the zero-width
quasiparticle results using Eq.~(\ref{chi_zero_width}) for fixed
$c$-quark mass, increasing in steps of 0.1~GeV from top to bottom. }
\label{fig_chi}
\end{figure*}
Let us end this section by a schematic evaluation of the widely
discussed ratio of viscosity to entropy density, $\eta/s$. Using
kinetic theory, one can roughly relate this quantity to
the spatial HQ diffusion coefficient as~\cite{Rapp:2008tf}
\begin{equation}
\frac{\eta}{s} \approx \frac{1}{5} T D_s \ .
\label{eta-s}
\end{equation}
The numerical coefficient probably constitutes a lower limit,
applicable to a weakly coupled system; it is most likely larger for
strongly coupled liquids, e.g.~$\sim$1/2 in AdS/CFT.
Nevertheless, the results for $U$ suggest that ${\eta}/{s}$
is not far from the conjectured lower limit of 0.08 for
quantum liquids. Moreover, its $T$-dependence seems to indicate a
minimum value when approaching $T_c$ from above which would resemble 
rather generic behavior of substances in the vicinity of a
critical point. This feature is also present for the $F$-potential,
albeit the  $T$-dependence is less pronounced; of course, the values
are also much larger compared to the $U$-potential, by about a factor
of 2-4. In fact, the ${\eta}/{s}$ value for $F$ at 2\,$T_c$ is
rather close to perturbative
calculations~\cite{Blaizot:1999xk,Arnold:2003zc} where $\eta/s\simeq 1$
with little $T$-dependence, indicating that resummation effects in
the $T$-matrix do not play a large role under these conditions.

\subsection{Charm-quark number susceptibility}
\label{ssec_chi}
Quark-number susceptibilities, $\chi_q$, which we already utilized 
in connection with the $c$-quark diffusion coefficient, 
Eq.~(\ref{gamma_c}), can be computed rather accurately in 
lQCD~\cite{Petreczky:2009cr}
and are therefore of great interest to constrain effective models of 
the QGP. For example, HQ number susceptibilities have been used to 
extract effective in-medium charm- and bottom-quark mass corrections by
fitting a zero-width quasiparticle expression, 
Eq.~(\ref{chi_zero_width}), to the lQCD results~\cite{Petreczky:2008px}.   
Here, we carry out a full off-shell calculation including finite-width
effects through the single $c$-quark propagators figuring into the 
charmonium spectral function~\cite{Hatsuda:1994pi},
\begin{eqnarray}
 \chi_{c}(T)=\frac{1}{T}\int\limits_0^\infty \frac{dE}{2\pi}
\frac{2}{1-\exp(-E/T)}\rho_{00}(E,\vec{0})\ .
\end{eqnarray}
For high $T\gg m_Q, \Sigma_Q$, this quantity 
reduces to 
\begin{eqnarray}
 \chi_{c}(T)=\frac{2\,N_c}{6}\,T^2 \ .
\label{chi_zero_mass}
\end{eqnarray}

In Fig.~\ref{fig_chi} we summarize our results for the HQ susceptibility
and compare to lQCD data as well as the zero-width quasiparticle
limit, Eq.~(\ref{chi_zero_width}), used in previous estimates. When 
plotted as a function of temperature over in-medium $c$-quark mass (left 
panel), the results for the $U$-potential are slightly above the 
ones for $F$ (note that the in-medium $c$-quark mass is 10-20\% larger 
for the $U$-potential, mostly due to the larger mass correction resulting 
from $U_\infty/2$ compared to $F_\infty/2$). The full results using $U$ 
are significantly above an off-shell calculation where the imaginary
parts are put to zero, which in turn agree very well with the zero-width
quasiparticle limit (dashed line).  
When plotted as a function of temperature (right panel in 
Fig.~\ref{fig_chi}), it turns out that the results for $F$ are
somewhat above those for $U$, due to the smaller $m_c(T)$. On the
other hand, the finite-width effects in the full results using $U$
produce an increase in $\chi_c$ over the no-width limit which 
corresponds to a $c$-quark mass decrease of ca.\,150-200\,MeV
in the zero-width quasiparticle expression. Thus, finite widths
considerably affect the extraction of the in-medium $c$-quark 
mass from the susceptibility. This effect also leads to a significant
improvement in the comparison to lQCD results: the full results with
$U$ roughly lie in the uncertainty band encompassed by the lQCD
results while the full results with $F$ tend to lie at the upper end
of that band. At the lowest considered temperature, the results for
$U$ seem to lie below the $N_t$=8 lQCD computations, but as indicated 
in connection with the large-$\tau$ limit of the $S$-wave charmonium 
correlator ratios, we believe that further width (and coupled-channel) 
corrections need to be included close to $T_c$ before more precise 
conclusions can be drawn. 
At least the underestimate of the lQCD data in our calculations for 
$\chi_c$ and the large-$\tau$ limit of $R_{PS,V}$ is consistent.   


\section{Conclusions}
\label{sec_concl}
%
We have conducted a study of charmonium and open-charm properties 
in the Quark-Gluon Plasma using a thermodynamic $T$-Matrix 
formalism\footnote{We 
recall that the static approximation encoded in the 3-D reduction 
schemes of the $T$-matrix can be justified since the energy-transfer 
for heavy-quark interaction is parametrically suppressed by $1/m_Q$.}. 
For the first time in the context of heavy quarks in the QGP, we have 
implemented this scheme selfconsistently at the one- and 
two-body level (i.e., selfenergy and scattering amplitude), including 
microscopically calculated off-shell effects, in particular imaginary
parts. Within the the Matsubara formalism, the two-particle 
propagator in the $T$-matrix equation automatically generates 
``zero-mode" contributions, i.e., scattering off pre-existing charm 
quarks in the heat bath. Following our earlier work~\cite{Riek:2010fk}
the two-body input potential was constructed using a field-theoretical 
model for Coulomb and confining forces with its 4 parameters fitted to 
finite-temperature lattice QCD data of the color-averaged heavy-quark 
free energy. As limiting cases of the interaction strength we have 
considered the resulting free and internal energy as an underlying 
potential. Once the input potential and the bare charm-quark mass are 
fixed (the latter is adjusted to reproduce the charmonium ground-state
mass in vacuum), there are no further tunable parameters in the 
heavy-heavy and heavy-light sector of our approach; heavy-quark 
interactions with gluons are treated perturbatively but turn
out to play a minor role (even with $\alpha_s$=0.4).
We are then able to comprehensively compute hadronic spectral functions 
in different charmonium and $D$-meson channels, as well as heavy-quark 
selfenergies and transport properties in the QGP. Specifically, we have 
applied these quantities to calculate euclidean correlator ratios for 
charmonia (consistently including zero-modes in scalar and axial/vector 
channels) and the charm-quark number susceptibility, which have both 
been ``measured" with good accuracy in thermal lattice QCD. 
Overall, we find an encouraging degree of agreement of our results 
with those in lQCD, especially at temperatures $T\ge1.5\,T_c$.
(possibly with a preference for using $U$ as potential), keeping
in mind that neither $F$ nor $U$, as computed in lQCD, necessarily 
provide a rigorous definition of a two-body potential, 
as both quantities are computed from differences of thermal averages. 
Especially close to $T_c$ a more complete description of charm and 
charmonia remains a challenge, requiring further refinements such 
as the inclusion of (pre-) hadronic states in the $QQ$ and $Qq$ 
$T$-matrices (coupled channels), a nonperturbative treatment of 
interactions with thermal gluons and a careful assessment of 
retardation effects. The large entropy contribution in the HQ
free energy close to $T_c$ indeed suggests many additional states 
to play a role. Another extension of our approach concerns the 
3-momentum dependence of charmonium 
correlators/spectral functions, for which interesting lQCD 
results are now becoming available~\cite{priv:2010}. 

Our analyses and findings reveal the intricate relation between 
charmonium bound-state properties and heavy-quark diffusion in the 
QGP, which, in particular, cannot be captured by perturbative 
treatments. The very same (resummed) force that generates resonance-like 
correlations (or threshold enhancements) in charmonium spectral 
functions -- crucial for properly describing the lQCD correlator ratios 
-- is operative in low-momentum charm-quark scattering (closely related 
to zero modes), and thus most likely instrumental in reducing the 
pertinent thermal relaxation times as required in phenomenological
applications to RHIC data~\cite{vanHees:2005wb,vanHees:2007me}. It 
remains an open question how large $c$-quark momenta need to be for 
radiative processes to become competitive with elastic 
scattering~\cite{Vitev:2007jj,Rapp:2009my}. 

In conclusion, we believe that studying the many-body physics of 
heavy-quark systems in the QGP can indeed reveal valuable insights 
into the medium modifications of the basic QCD force via its
effects on bound-state properties and heavy-flavor transport.
The nonperturbative nature of the problem, especially in the vicinity
of $T_c$, requires effective approaches whose predictive power 
greatly benefits from constraints obtained from thermal lattice QCD.  
This will hopefully pave the way toward illuminating and quantifying
the properties of the QGP even in the case of rather strong coupling,
and help understand some of the fascinating phenomena observed in
ultrarelativistic heavy-ion collisions. 

\vspace{0.5cm}

{\bf Acknowledgments}\\
The authors acknowledge useful discussions with M.~He and X.~Zhao,
and are indebted to J.I.~Skullerud for providing the lattice-QCD data 
on the charmonium correlator ratios.
This work has been supported by the U.S. NSF under CAREER grant 
No.~PHY-0449489 and grant No.~PHY-0969394, and by the 
A.-v.-Humboldt Foundation. 
%
%
%
%
%
%
%
%
%
%

%
\bibliographystyle{prsty}
\end{document}